\documentclass[aps,prd,preprint,showpacs,amsmath,amssymb]{revtex4}
\usepackage{epsfig}

\newlength{\www}

\newcommand{\be}{\begin{equation}}
\newcommand{\ee}{\end{equation}}
\newcommand{\ba}{\begin{eqnarray}}
\newcommand{\ea}{\end{eqnarray}}

\begin{document}

\title{\vspace{1cm}
A complete one-loop description of associated $tW$ production at LHC and 
a search for possible genuine supersymmetric effects
}
\author{
M.~Beccaria$^{a,b}$, 
C.M.~Carloni Calame$^{c}$, 
G.~Macorini$^{d, e}$,
G.~Montagna$^{f, g}$, 
F.~Piccinini$^{g, f}$, 
F.M.~Renard$^h$ 
and C.~Verzegnassi$^{d, e}$ \\
\vspace{0.4cm}
}

\affiliation{\small
$^a$ $\mbox{Dipartimento di Fisica, Universit\`a del Salento, Italy}$ \\
\vspace{0.2cm}
$^b$ INFN, Sezione di Lecce, Italy\\
\vspace{0.2cm}
$^c$ $\mbox{School of Physics \& Astronomy, Southampton University, UK}$ \\
\vspace{0.2cm}
$^d$
$\mbox{Dipartimento di Fisica Teorica, Universit\`a di Trieste, Italy}$ \\
\vspace{0.2cm}
$^e$ INFN, Sezione di Trieste, Italy\\
\vspace{0.2cm}
$^f$ $\mbox{Dipartimento di Fisica Nucleare e Teorica, Universit\`a di Pavia, Italy}$ \\
\vspace{0.2cm}
$^g$INFN Sezione di Pavia, Italy\\
\vspace{0.2cm}
$^h$ Laboratoire de Physique Th\'{e}orique et Astroparticules, 
Universit\'{e} Montpellier II, France
}

\begin{abstract}
We compute, in the MSSM framework, the sum of the one-loop electroweak 
and of the total QED radiation effects for the process $pp \to t W+X$,
initiated by the parton process $bg\to tW$. Combining these terms with the
existing NLO calculations of SM and SUSY QCD corrections, we analyze 
the overall one-loop supersymmetric effects on the partial rates of the process, obtained
by integrating the differential cross section up to a final variable
invariant mass.
We conclude that, for some choices of the SUSY parameters and for
relatively small final invariant masses, they could become relevant under
realistic experimental conditions at LHC.
\end{abstract}


\begin{flushright}
PTA/07-17\\
FNT/T 2007/02
\end{flushright}

\maketitle

\section{Introduction}
\label{sec:intro}

Single top production at LHC has already been emphasized by several
authors~\cite{wagner} as one of the processes to be possibly
measured with maximal experimental accuracy. The main and known reason is
the fact that in the Standard Model from this measurement one would be
able to derive the first
direct determination of the $Wtb$ coupling $V_{tb}$, assumed to be practically
equal to one if unitarity of the CKM matrix is postulated. Since the rates
of the three different single top  production processes, usually called
$t$-channel, $s$-channel, 
and associated production, are proportional to $|V_{tb}|^2$, 
an accurate measurement of those
rates would correspond, in principle, to a correspondingly accurate
measurement of $V_{tb}$. From the preliminary experimental 
analyses~\cite{yreport:1999} 
one hopes that a realistic overall accuracy might 
eventually reach the ten percent size. The corresponding precision of
the determination of $V_{tb}$ would be therefore of five percent, if
no extra theoretical uncertainties had to be added.

The previous statement requires a description of the status of the
existing theoretical calculations. A complete NLO QCD calculation exists
for all the three single top processes in the Standard 
Model~\cite{qcdsingletop} 
 with an attached uncertainty that varies with the process. 
It seems probable that this uncertainty will be essentially reduced by the
future LHC related measurements, and we shall return to this point later on
in the paper. For the three production channels, SUSY NLO QCD
effects have been recently evaluated in the MSSM~\cite{Zhang:2006cx}. 
The pure electroweak one-loop effects in the MSSM have 
been computed recently for 
the associated $tW$ production case~\cite{Beccaria:2006dt}. 
This calculation did not contain, though,
the QED radiation corrections. In the case of the total rate, a modest (few percent)
relative effect was found. In a subsequent note~\cite{yellow} 
it was shown that, still at the pure
electroweak one-loop level, more interesting effects could be found
from the measurement of partial rates, defined as an
integral of the differential cross section performed from threshold to a
final suitable invariant mass. In particular, for low values of the
latter, a positive effect was computed, that 
obviously could not be due to a violation of the CKM matrix structure leading to smaller values of $V_{tb}$.

Starting from these premises, the purpose of this paper is that of
providing a really ``complete'' calculation of one-loop effects in the MSSM
on the partial rates of the $tW$ production process, where for each partial
rate the QED effect is also included. Moreover, we shall provide a
preliminary evaluation of the overall
theoretical uncertainties taking into account the variation of the final
considered invariant mass. The result will be, essentially, an indication
of an ``optimal'' choice of partial rate, {\it i.e.} of one for which the two
combined requests of maximal electroweak effect and of minimal overall
uncertainty might be satisfied. This could lead to a relevant precision
test of the MSSM at LHC, whose accuracy would certainly improve with time
if preliminary encouraging  signals were revealed, motivating more
accurate dedicated experimental measurements.

The paper will be organized in the following way: 
section 2 will contain a review of the electroweak calculation,
essentially shortened since all the necessary details have 
already been given in Ref.~\cite{Beccaria:2006dt}. In section 3, 
the new calculation of the additional QED radiation
effect will be described. Section 4 will contain a short discussion of the
existing QCD calculations and of their uncertainties, taking into account the
considered variation of the final invariant mass. 
Summarized results will be discussed in Section 5 
and a few conclusions will finally appear in Section 6.

\section{Electroweak one-loop effects}
\label{section:electroweak}

\begin{figure}
\centering
\epsfig{file=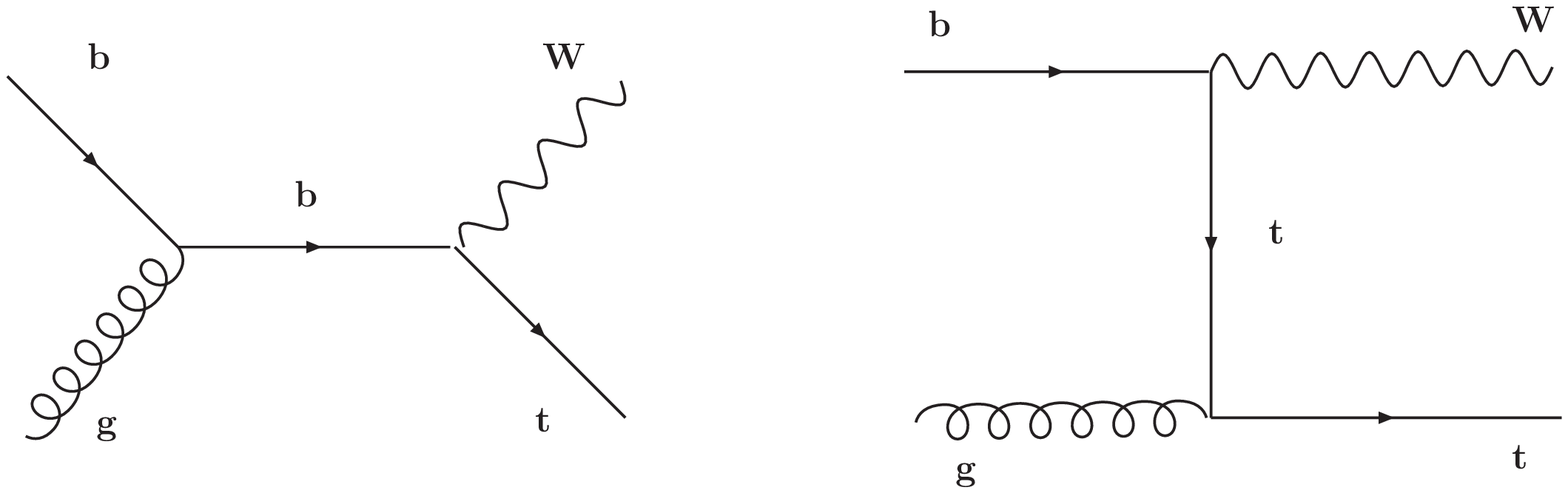, width=12cm, angle=0}
\caption{
Born diagrams for the process $bg\to tW^-$.}
\label{fig:born}
\end{figure}

At the partonic level the associated $tW^-$ production is initiated by 
the process $bg \to tW $, described at Born level by the two diagrams in 
Figure~\ref{fig:born}.

In a previous paper~\cite{Beccaria:2006dt}, 
we have analyzed the complete one-loop electroweak corrections
to this process treating QED effects in the soft-photon approximation. 
Here, we summarize briefly the results of Ref.~\cite{Beccaria:2006dt}
as a background to the complete analysis including hard QED radiation to be 
discussed in the remaining part of the present paper.

Our starting observable for this process was the invariant mass 
distribution defined as 
\ba
\label{eq:basic}
\frac{d\sigma(pp \to t W^-+X)}{dM_{tW}} &=& \int \, 
dx_1 \, dx_2 \, d\cos\theta \,[~ b(x_1, \mu) g(x_2, \mu)+g(x_1, \mu)
b(x_2, \mu)~] \nonumber \\ 
&\times & \frac{d\sigma_{bg\to  t W^-}}{d\cos\theta}
\, \delta(\sqrt{x_1 x_2 S} - M_{tW}) \, ,
\ea
\noindent
where $\sqrt{S}$ is the proton-proton c.m. energy, $M_{tW}$ 
is the $tW$ invariant mass, $\theta$ is the top-quark scattering angle 
in the partonic c.m. frame, and $i(x_i, \mu)$ are the distributions 
of the parton $i$ inside the proton
with a momentum fraction $x_i$ at the scale $\mu$.

The invariant mass distribution $d\sigma/dM_{tW}$ has been evaluated for 
a number of SUSY benchmark points
with a wide variation of mass spectra and values of the mixing parameter 
$\tan\beta$. In particular, the considered MSSM points have been 
the standard ATLAS DC2 SU1 and SU6~\cite{DC2} 
and two generic mSUGRA  points with light spectrum. The calculation has 
been performed with a kinematical 
cut on the $W$ or top transverse momentum
$p_{T, \min} = 15$ GeV, and we have included a QED soft-photon 
contribution, computed assuming an upper value of the soft-photon 
energy $\Delta E = 0.1$~GeV.
The main results of our detailed numerical investigation can be summarized
as follows. The electroweak one-loop effect in the MSSM is the sum of the
pure SM and of the genuine SUSY components. These two terms have a
fundamentally different dependence on the invariant mass of the final
state. More precisely, the SM effect varies from positive values of
approximately 5\% in the lowest invariant mass
 range to larger negative
values in the high invariant mass sector.
 One actually expects from general
considerations  such  negative effects, coming from a logarithmic
contribution of Sudakov kind to the asymptotic value of the scattering
amplitude. As a consequence of the change of sign of the SM contribution,
the overall effect in the total rate is practically vanishing.
The genuine SUSY effect has a rather different nature. It remains
systematically positive in all the invariant mass range (realistically
considered up to a final value of 1~TeV) for all the considered SUSY
benchmark points, assuming essentially the same relative effect of 3-4~\% 
in all cases, with
a maximum value reached in the SU6 point. The lack of a negative large
energy effect can be understood as a consequence of the fact that in the
considered benchmark points there are relatively large SUSY masses that
appear in the virtual loops, which hides the appearance of asymptotic
logarithmic Sudakov effects.
A possibility that was considered in the note~\cite{yellow} 
is that of considering partial rates, in particular low energy
partial rates, obtained by integrating the differential cross section from
threshold to a given final invariant mass. If the latter is fixed at small but
meaningful values (say, 400~GeV), the relative SM and genuine SUSY
effects sum up. Also, the effect will be positive, therefore not
possibly due to those violations of the CKM matrix structure which would decrease the
value of $V_{tb}$. This idea is confirmed by the detailed numerical
calculation, shown in the next Figure~\ref{fig:twyellow} 
taken from Ref.~\cite{yellow} 
and computed with the same kinematical 
and infrared cuts as in Ref.~\cite{Beccaria:2006dt}. 
One sees that for values of the final invariant mass in the
range mentioned above
the MSSM electroweak effect is not negligible, reaching
values of 6-7\%, that appear possibly relevant.

The previous considerations miss two important points. The first one is an
original  calculation
of the complete (soft and hard) QED effect, which has never been performed
before. The second one is a summary of the QCD NLO calculations, including
both SM and SUSY QCD (the latter one will be in fact rather useful  for
our purposes). These two topics will be discussed in the forthcoming
sections.

\section{QED radiation}
\label{section:qed}
The ${\cal O}(\alpha)$ electroweak corrections include contributions 
from virtual and from real photon emission. The virtual photon exchange 
diagrams belong to the complete set of electroweak virtual corrections, 
and are necessary for the gauge invariance of the final result. 
The singularities associated with the massless nature of the photon have 
been regularized by introducing a small photon mass $m_\gamma$. 
The real radiation contribution has been split into a soft part, 
derived within the eikonal approximation, where the photon energy has 
been integrated from the lower bound $m_\gamma$ to a maximum cut off 
$\Delta E$, and into a hard part, evaluated 
by means of a complete calculation of the diagrams shown in 
Figure~\ref{fig:QEDdiagrams}. The soft real contribution 
contains explicitly the photon mass parameter $m_\gamma$. 
The logarithmic terms containing $m_\gamma$ cancel 
exactly in the sum of virtual and soft real part, leaving only 
polynomial spurious terms, which approach zero at least as $m_\gamma^2$. 
The hard contribution, integrated from
the minimum photon energy $\Delta E$ to the maximum 
allowed kinematical value, 
can be calculated with a massless external photon. The complete  
matrix element for real radiation, including fermion mass effects, 
has been calculated analytically with the help of FORM~\cite{form}, in order 
to handle efficiently the traces of strings of Dirac gamma matrices. 
As an internal check, the complete matrix 
element has been verified to recover (in the limit $k \to 0$, where 
$k$ stands for the photon momentum) the analytical expression 
of the eikonal approximation, 
factorized over the tree-level amplitude.

The integration over the phase space has been performed numerically 
with Monte Carlo methods. In order to treat efficiently the regions 
related to the infrared and collinear singularities, the importance 
sampling technique has been adopted. In particular the photon energy 
is generated in the partonic center of mass system according to the 
distribution $1 / E_\gamma$; the photonic angular variables are generated 
with a multichannel strategy according to the distributions 
$1 / (1 - \beta_i \cos\vartheta_{\gamma i})$, where $i = t, b, W$,  
$\beta_i$ represent their velocities, and 
$\vartheta_{\gamma i}$ are the relative angle between the photon and 
the charged particles $i$. 
 
The final cross section has to be independent of the 
fictitious separator $\Delta E$, for sufficiently small $\Delta E$ values. 
This has been checked numerically to hold at the level of few 0.01\% 
for $\Delta E \leq 1$~MeV, as shown in Figure~\ref{fig:QEDcheck} 
(right panel), 
despite the strong sensitivity to $\Delta E$ of the soft 
plus virtual and of the hard 
cross section separately, as shown in Figure~\ref{fig:QEDcheck} (left panel). 
The leading logarithmic dependence on $\Delta E$ 
(with opposite coefficients) in the soft plus virtual 
and in the hard cross sections
is manifest in the logarithmic scale plot. 

\section{QCD effects}
\label{section:qcd}
\subsection{NLO SM corrections}
\label{subsection:nlo}
The NLO SM QCD corrections to the $tW$ signature 
have been calculated by various authors~\cite{qcd1,qcd2,qcd3} 
within the on-shell approximations for $W$ boson and top quark. In 
Ref.~\cite{qcd3} the corrections considering also the decay of the top quark 
have been evaluated. Such a calculation is implemented in the fixed-order 
Monte Carlo program MCFM~\cite{mcfm}, v5.1, which we use to estimate 
the QCD uncertainties associated with the integrated $t W$ mass distribution. 
In our simulation we adopt the on-shell approximation for $W$ and top quark 
(available as an option in the MCFM code), consistently with the 
electroweak calculation presented in this study. The input parameter 
values of the program have been tuned with the ones adopted in the 
electroweak calculation. For internal perturbative consistency we 
used in the NLO calculation the CTEQ6M set for PDFs. 
In addition, at NLO there is the need to fix the 
factorization and renormalization scales, $\mu_F$ and $\mu_R$, respectively. 
The typical way of studying the remaining QCD theoretical 
uncertainty of the predictions, as due to missing higher order terms, is 
to vary $\mu_F$ and $\mu_R$ around the relevant scale of the process, 
which, in the case under study, would be given by $m_t + m_W$. However, 
the presence of a $b$ quark in the initial state introduces two subtleties, 
which have been addressed in Ref.~\cite{qcd3}. The real ${\cal O}(\alpha_s)$ 
radiation contribution contains diagrams where an initial state gluon 
splits into a $b \bar b$ pair giving rise to the $W t b$ final state. 
The collinear $g \to b \bar b$ splitting is already accounted for in the 
$b$-quark distribution function, used in the lowest order calculation. 
Therefore the net contribution from the $g g \to W t b$ diagrams should 
be approximately zero, including appropriate counter-terms and 
integrating over all $b$-quark transverse momenta up to $\mu_F$. In 
Ref.~\cite{qcd3} this has been checked to happen for $\mu_F \leq 65$~GeV. 
An additional problem associated with the $g g \to W t b$ diagrams arises 
in the portion of the phase space where the $W b$ system crosses 
over the pole of the virtual top-quark propagator. Actually this contribution 
represents the doubly resonant $t \bar t$ production process, and it 
is preferable to exclude it from the NLO corrections to the $t W$ 
process~\cite{qcd3}. This is achieved in MCFM by applying a veto on the 
$p_T$ of the additional $b$ quark that appears at next-to-leading order. 
For consistency, the maximum allowed $p_T$ of the $b$ quark should 
be chosen of the same order of $\mu_F$. 

For the above reasons we have selected $\mu = \mu_F = \mu_R = 50$~GeV and 
$p_T^{{\rm b}\, \, {\rm veto}} = 50$~GeV, as in Ref.~\cite{qcd3}, 
and studied the variation of the resulting $K$-factor (defined as 
$\sigma_{NLO} / \sigma_{LO}$) in the range 
25~GeV $\leq \mu \leq $ 255~GeV. The $K-$factor ranges from 1.26 for 
$\mu = 25$~GeV to 1.12 for $\mu = 50$ and $100$~GeV 
(1.17 for $\mu = 255$~GeV). The inclusive NLO cross section varies from 
36.07(1) pb for $\mu = 25$~GeV to 34.84(1)~pb for $\mu = 50$~GeV,  
35.09(1)~pb for $\mu = 100$~GeV and 35.86(1)~pb for $\mu = 255$~GeV, 
thus showing a stability at the level of 
few per cent, in agreement with Figure 7 of Ref.~\cite{qcd3}. 

At this point we can study the stability of the QCD NLO predictions 
on the partial rates. In Figure~\ref{fig:QCDKfact}, 
we quantify the size of the NLO QCD corrections showing 
the differential $K$-factor, with $p_T^{{\rm b}\, \, {\rm veto}} = 50$~GeV.
The curves correspond to the values $\mu = 25$ (solid line), 
$50$ (dot-dashed line) and $100$ GeV (dashed line) in the NLO calculation, 
while the scale of LO calculation is kept fixed at the value $\mu_0 = 50$~GeV. 
The size of the QCD corrections is of the order of 20\%, decreasing by 
about 10\% when $M_{tW}$ ranges from threshold to 1 TeV. 
The scale uncertainties are lower than about 4\%, 
being maximal at the highest $M_{tW}$. 

\subsection{SUSY QCD corrections}
\label{subsection:susyqcd}

The one-loop SUSY QCD effects to the three channels of single top production have been computed at LHC
in Ref.~\cite{Zhang:2006cx}. These are radiative corrections with propagation of virtual gluinos
in the quantum loop. The numerical analysis of 
Ref.~\cite{Zhang:2006cx} is performed 
in the constrained MSSM within mSUGRA. The MSSM parameters are the five inputs at grand unification 
scale
\be
M_{1/2},\  M_0,\  A_0,\  \tan\beta, \ \mbox{sign}\, \mu,
\ee
where $M_{1/2}, M_0, A_0$ are the universal gaugino mass, scalar mass, and the
trilinear soft breaking parameter in the superpotential. This last parameter
has been set to $A_0=-200~\mathrm{GeV}$. The sign of $\mu$ is positive.

The effects in the associated production channel are studied by evaluating the $K$ factor 
defined as the ratio of the SUSY QCD corrected cross
sections to LO total cross sections, calculated using the CTEQ6M
PDFs set. The dependence of the $K$ factor on the various MSSM parameters is 
analyzed in great detail.

The dependence on the gluino mass $M_{\tilde{g}}$ ($M_{1/2}$)  can be studied at various 
values of $\tan\beta=5,20$ and $35$. 
The K factor increases with  $M_{\tilde{g}}$ for small
$M_{\tilde{g}}(\lesssim 160\mathrm{GeV})$, while it decreases
 with $M_{\tilde{g}}$ for large $M_{\tilde{g}}(\gtrsim
160\mathrm{GeV})$. 
In general the dependence on $\tan\beta$ is rather mild.
The typical values are about $K=1.06$.

For example, assuming $\tan\beta=5$ and computing 
the $K$ factor as functions of
$M_{\tilde{g}}$($M_{1/2}$) for different $M_0$, one finds that
there are large variations in $K$ when $M_{\tilde{g}}$
becomes small. 
It decreases with $M_{\tilde{g}}$ and especially rapidly
when $M_{\tilde{g}}\lesssim150\mathrm{GeV}$
at least 
for $M_0=150\mathrm{GeV}$.
Nevertheless, as soon as $M_{\tilde{g}}$ becomes large,
the decoupling of heavy gluinos ($M_{\tilde{g}}\gtrsim450\mathrm{GeV}$)
gives saturated stable values of $K$. Again $K\simeq 1.06$
for $M_{\tilde{g}}\lesssim500\mathrm{GeV}$. 

Similar results are obtained by varying the stop mass $M_{\tilde{t}_1}$ ($M_0$), 
assuming $\tan\beta=5$, and $M_{1/2}=40, 70$ and $100\mathrm{GeV}$, respectively. 
The $K$ factor is about $1.06$ for most values 
of $M_{\tilde{t}_1}$, decreasing slowly with
$M_{\tilde{t}_1}$.

In conclusion, the typical SUSY QCD correction to the process of associated
production is about $+6\%$  for most values of the explored parameters values.

\subsection{PDF uncertainties}
\label{subsection:pdf}
Another source of theoretical uncertainty is given by the contribution 
of the parametric errors associated with the parton densities. This 
could become particularly relevant for single top channels, due to the 
presence of an initial state $b$ quark, 
whose distribution function is strictly related to the gluon distribution. 
We have studied the impact of such uncertainties by using 
the NLO PDF sets MRST2001E and CTEQ61 as in the LHAPDF 
package~\cite{lhapdf}. The results are shown in Figure~\ref{fig:pdfunc}.
The spread of the predictions obtained with the MRST set displays a 
relative deviation of about 1\%, while the CTEQ set gives a larger 
uncertainty reaching the 3\% level for high $t W$ mass. 
This is due to different values of the tolerance parameter~\cite{Tpdf}, the 
latter being defined as the allowed maximum of the $\Delta\chi^2$ variation 
w.r.t. the parameters of the best PDFs fit. 
Conservatively, we can associate to our predictions an uncertainty due 
to the present knowledge of parton densities of about 3\%. However it is worth 
noting that the uncertainties obtained according to such a procedure 
are of purely experimental origin only ({\it i.e.} as due to the systematic 
and statistical errors of the data used in the global fit), leaving aside 
other sources of uncertainty of theoretical origin.


\section{Results}

Coming back to the genuine electroweak corrections, we present in this 
section our results for the 
distribution $d\sigma/d M_{tW}$ and for the integrated cross section 
in the Standard Model and in the MSSM. The numerical results
shown in the following have been obtained in terms of the LO PDF set 
CTEQ6L with $\mu = m_t + M_W$, the latter being the top-quark and
W-boson mass, respectively.

We begin with the invariant mass distribution $d\sigma/dM_{tW}$. 
We show in Figure~\ref{fig:effdistsm}
the electroweak effect in the Standard Model case. 
The two lines allow to appreciate the effect of the hard photon radiation.
It adds a positive contribution to the effect that is uniformly 
positive over most of the explored energy range.
The effects in the MSSM are shown in  Figure~\ref{fig:effdistsusy} 
at the two benchmark points SU1 and SU6. The pattern is 
similar to that of the Standard Model, in agreement with the general partial 
decoupling of the genuine SUSY components. 

The behaviour of the integrated cross section is shown in  
Figure~\ref{fig:effintsm}, for the Standard Model, and in 
Figure~\ref{fig:effintsusy} for the MSSM. Again, 
the pattern is similar. The cross section is integrated from threshold up to 
a maximum invariant mass  $M_{tW}^{\rm max}$. 
The electroweak effect is always positive and is maximum for 
small $M_{tW}^{\rm max}$. 
This is a consequence of a coherent sum
of positive one-loop effects coming from the electroweak  sector of the
MSSM and from the complete QED contribution. For larger $tW$ invariant
masses, the electroweak SM contribution decreases, and the overall effect
is weakened. For what concerns the  electroweak SUSY effect, 
it remains of the order of
a few (positive) percent in all the considered benchmark points, the
largest effect being obtained in the SU6 case.
 This is not, though, the total genuine SUSY effect. In fact, to obtain
the latter, one still has to add the SUSY QCD contribution of 
Ref.~\cite{Zhang:2006cx}. The corresponding effect on the integrated 
partial rates is shown
by the dashed curves in the right panels of Figure~\ref{fig:effintsusy} 
in a qualitative, but essentially correct, way, simulating the SUSY QCD 
component by a constant 6\% shift, consistently with the analysis of 
Ref.~\cite{Zhang:2006cx}. From an inspection  of Figure~\ref{fig:effintsusy} 
one can conclude that, 
for final invariant masses of the 400~GeV size, 
an overall one loop effect of approximately
13-14\% is produced in the MSSM by the positive sum of electroweak
and SUSY QCD contributions. The size of the genuine one-loop SUSY effect
is of, roughly, ten percent. To obtain the complete value of the rates it
is now sufficient to add to the values of Figure~\ref{fig:effintsusy} 
the remaining SM QCD effect, exhaustively illustrated in section~4. 
In this way, the complete one-loop expression  of the rates can be obtained. 
In fact, from the calculations that we have performed, the values of 
other possibly interesting observable quantities can be easily obtained. 
The reason why we have limited our presentation to the partial 
rates will be summarized in the final conclusions.

\section{Conclusions}
\label{section:conclusions}
We have performed in this paper the first (to our knowledge) complete
calculation of the one-loop effect on the process of $tW$ production, 
including a discussion of the overall size of the theoretical uncertainties. 
Our interest has been concentrated on the particular quantities that we have
defined as partial rates, with special emphasis on the low (400~GeV) 
final invariant mass. The reason of this interest is actually twofold,
since with this choice the related quantity meets, at the same time,  two
conditions on the purely theoretical side. In fact, it maximizes  the 
overall one-loop electroweak (including QED) effect, that can reach the 
8\% size, a value that should not be neglected. With the addition of SUSY 
QCD one-loop terms, the genuine SUSY contribution reaches the 10\% size.
At the same time, it maximizes the theoretical uncertainties that we have considered in the paper.
On a purely experimental side, 
we do not have yet at our disposal a rigorous 
experimental analysis of the $tW$ production process, which seems to us extremely relevant and necessary. 
It is well known that the largest background for associated $tW$ production is top quark pair production.
From the point of view of signal identification, the region with small $tW$ invariant mass, near the $t\overline{t}$ threshold,  
has possibly a chance  to be optimal.
Waiting for a  dedicated effort we can though rely on the fact that 
a measurement of the rate at the ten percent level should 
be considered as a ``must'' project for the study of single top production. 
Should this result be met, the measurement of our
partial rates could indeed represent a relevant and original 
test of genuine SUSY effects in the MSSM at LHC.

\vskip 24pt\noindent
{\bf Acknowledgements}\\
We are grateful to Marina Cobal for discussions. 
The work of C.M. Carloni Calame is supported by a Royal Society
and British Council Short Visit grant.

\newpage

\hfill

\newpage

\begin{figure}
\centering
\epsfig{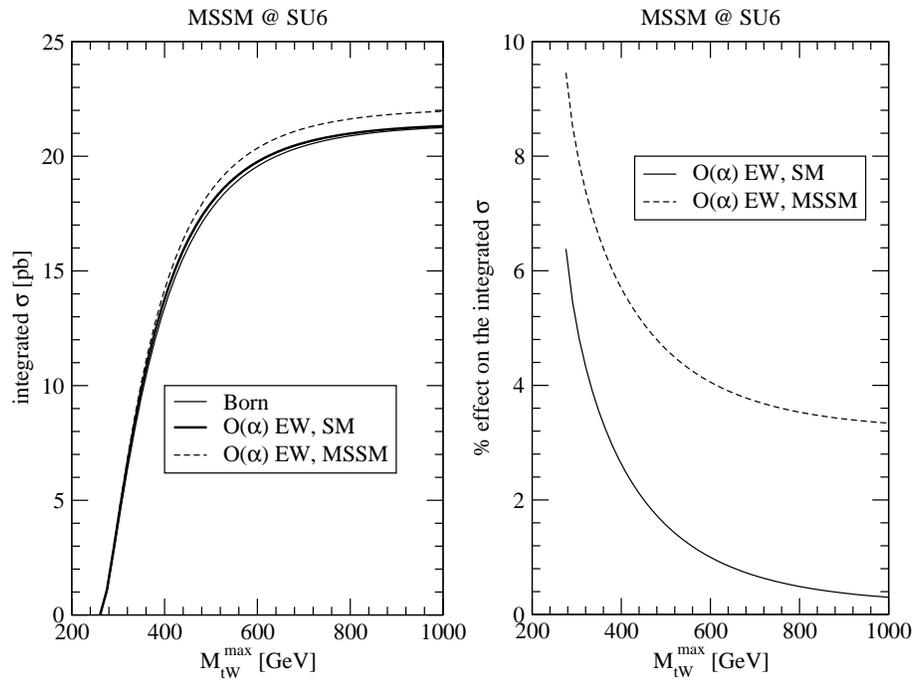}
\caption{Integrated cross section (from threshold up to $M_{tW}^{\rm max}$) in the soft-photon approximation.}
\label{fig:twyellow}
\end{figure}

\newpage

\begin{figure}
\centering
\epsfig{file=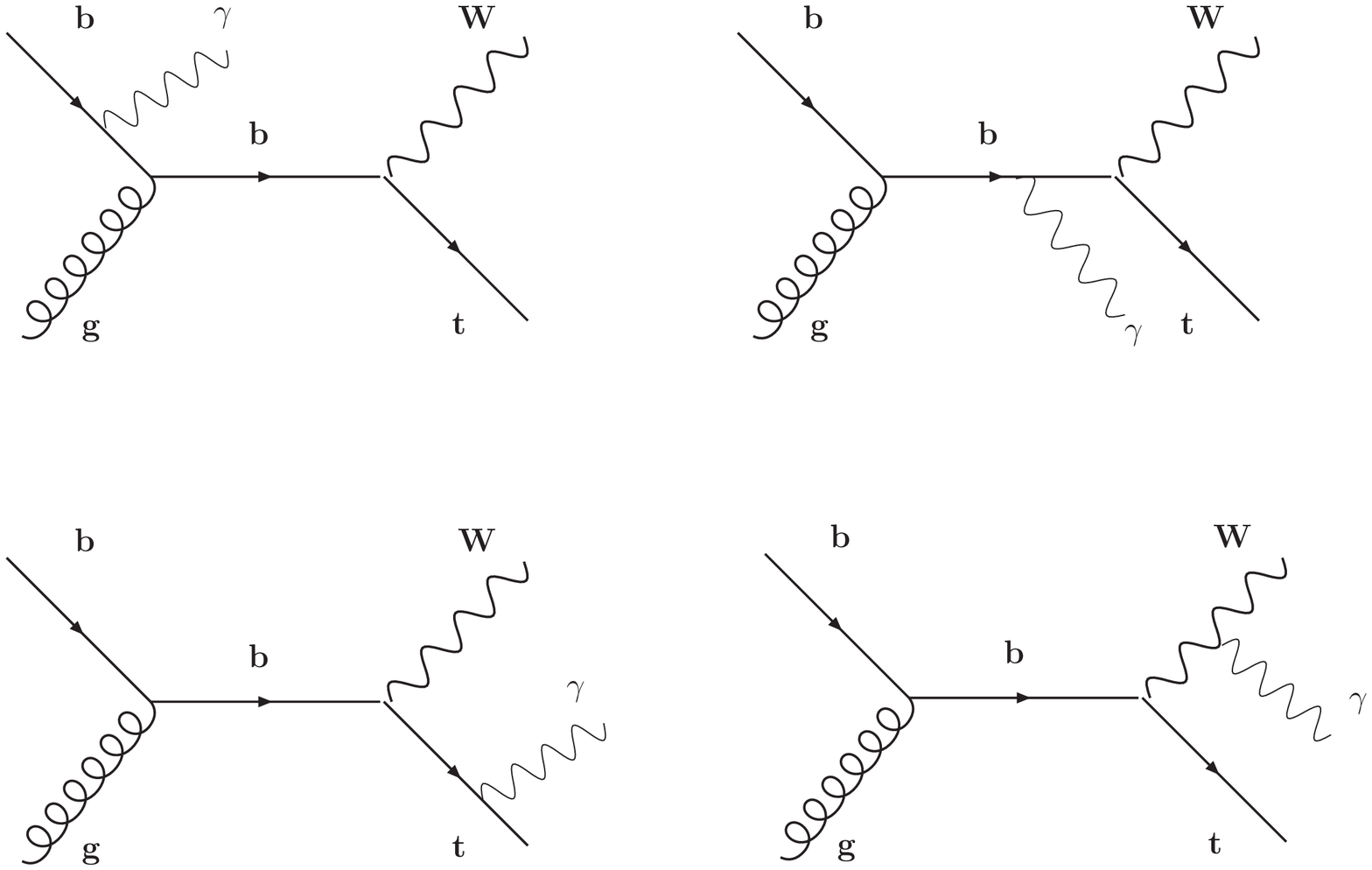, width=12cm, angle=0}
\vskip 1truecm
\epsfig{file=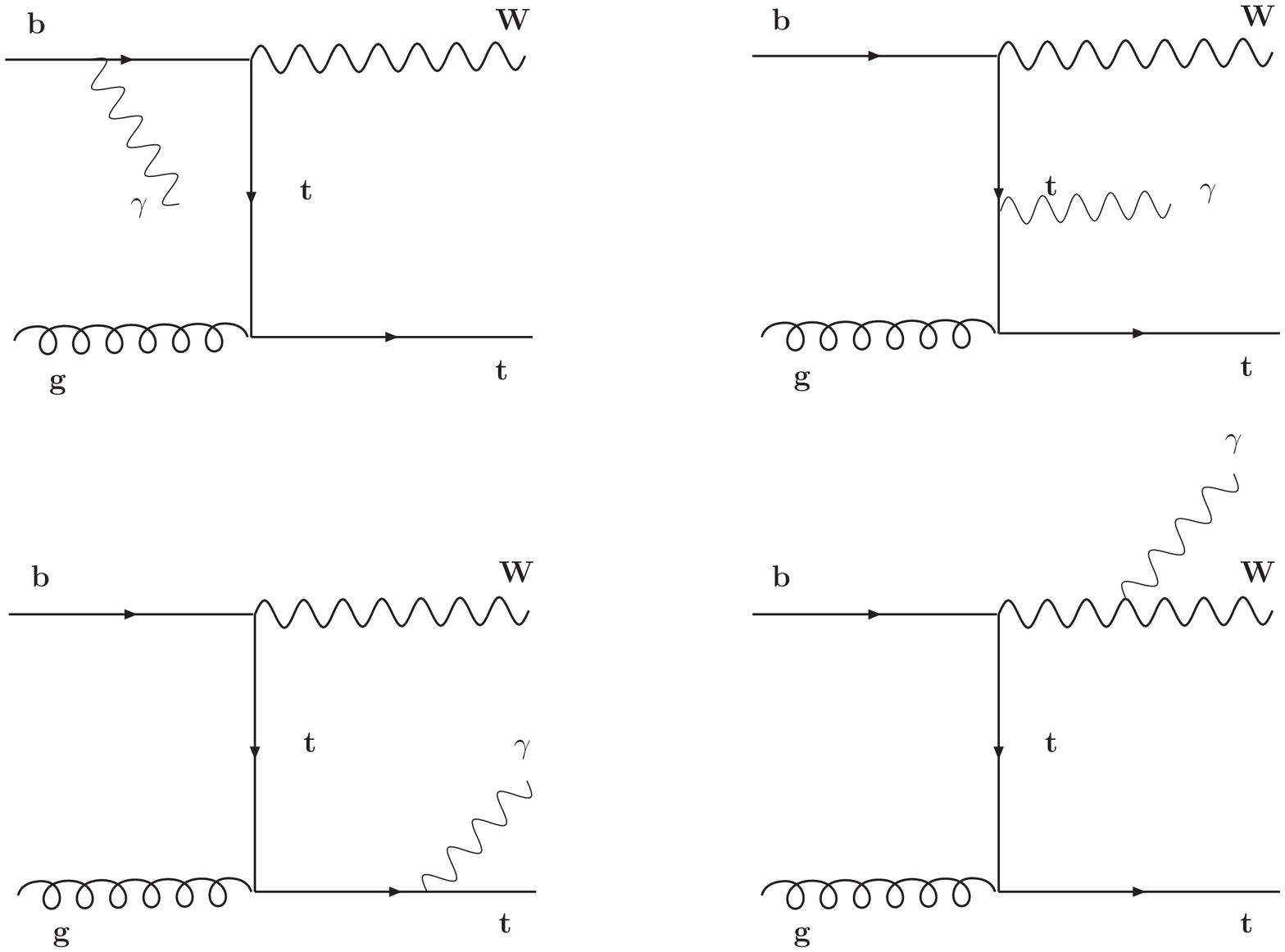, width=12cm, angle=0}
\caption{
Diagrams for the radiative process $bg\to t W^- \gamma$.}
\label{fig:QEDdiagrams}
\end{figure}

\newpage

\begin{figure}[htb]
\begin{center}
\epsfig{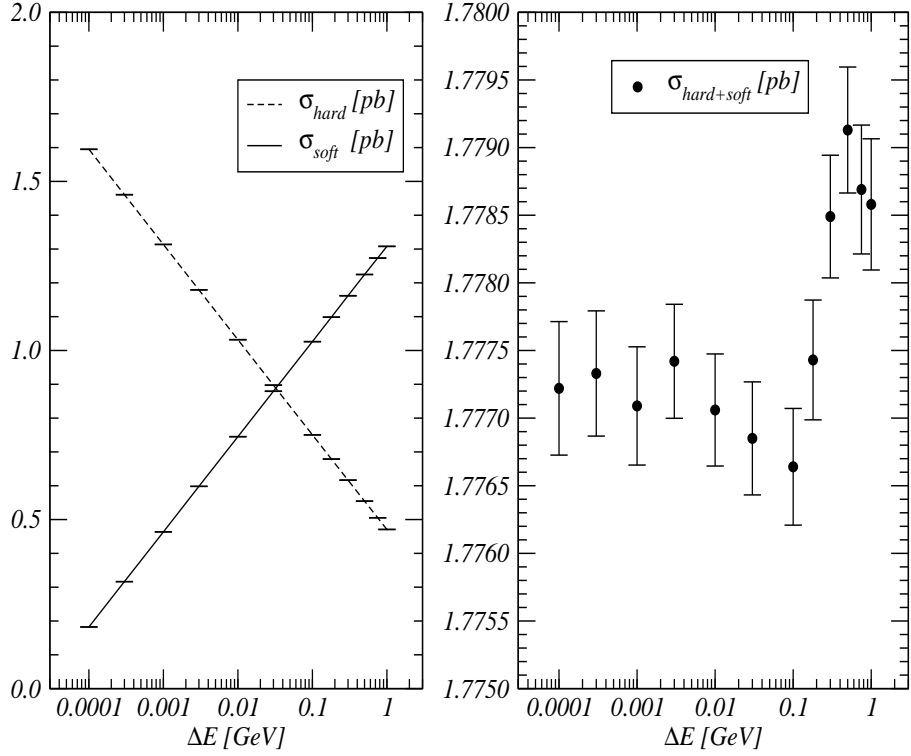}
\caption{\label{fig:QEDcheck} Left: dependence of the ${\cal O}(\alpha)$ 
soft plus virtual and hard cross sections on the soft-hard 
separator $\Delta E$. Right: independence of the sum of ${\cal O}(\alpha)$ soft plus 
virtual and hard cross sections 
of the separator $\Delta E$, checked numerically up to an accuracy 
of the order of $0.01\%$.}
\end{center}
\end{figure}

\newpage

\begin{figure}
\begin{center}
\epsfig{file=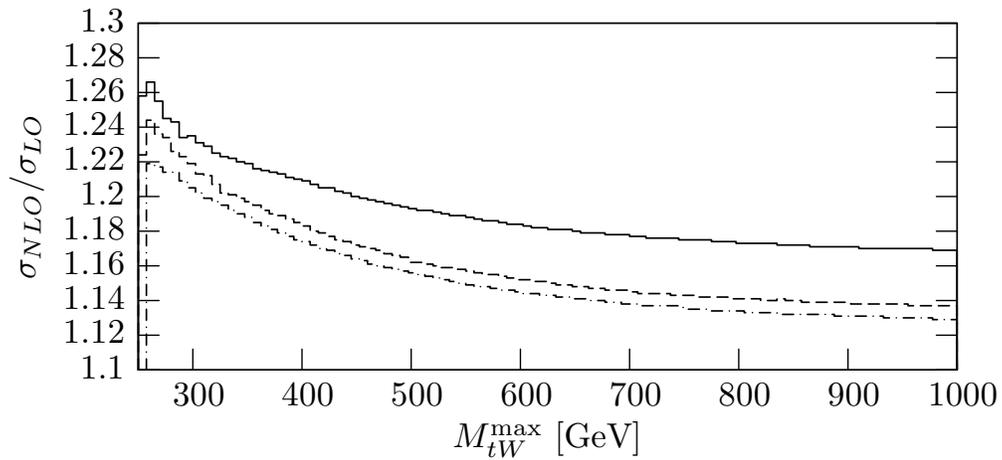,height=6.cm,angle=0}
\caption{\label{fig:QCDKfact} QCD $K$-factor 
($\sigma_{NLO}(\mu)/\sigma_{LO}(\mu_0=50\ {\rm GeV})$) as a function of 
$M_{t W}^{\rm max}$. Solid line, $\mu = 25$~GeV, dot-dashed line,  
$\mu = 50$~GeV, dashed line, $\mu = 100$~GeV. }
\end{center}
\end{figure}

\newpage

\begin{figure}
\begin{center}
\epsfig{file=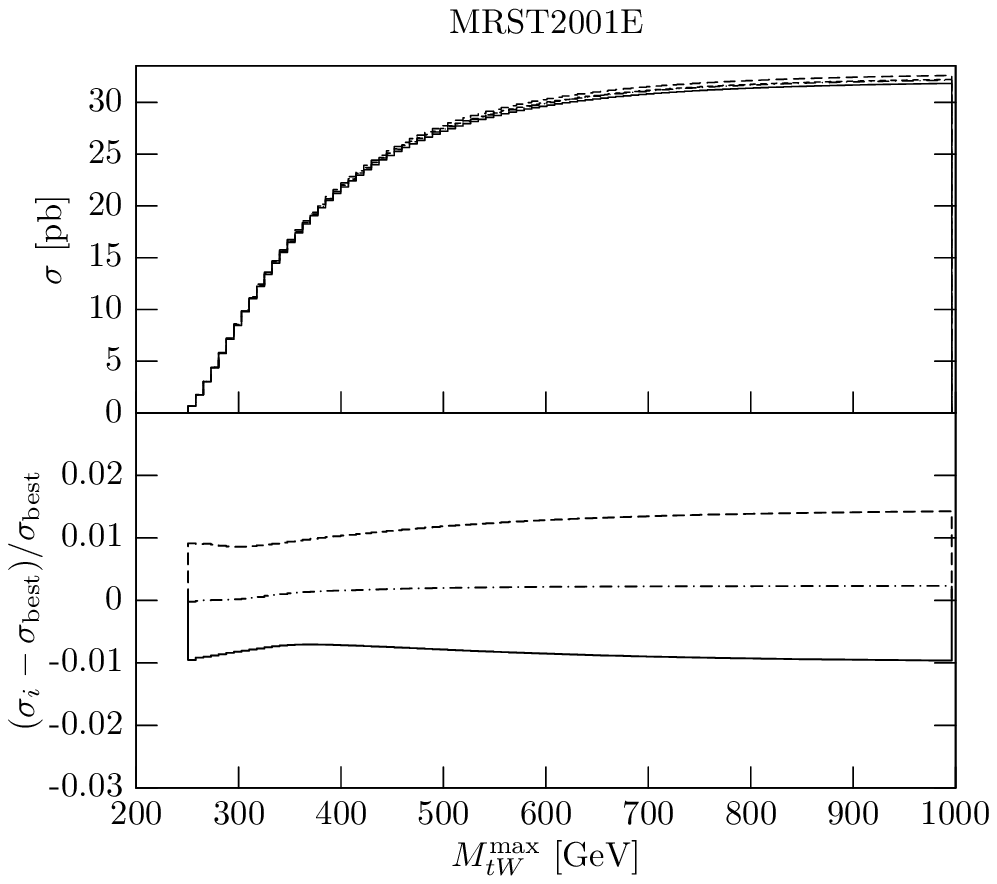,height=8.5cm,width=8cm,angle=0}
\epsfig{file=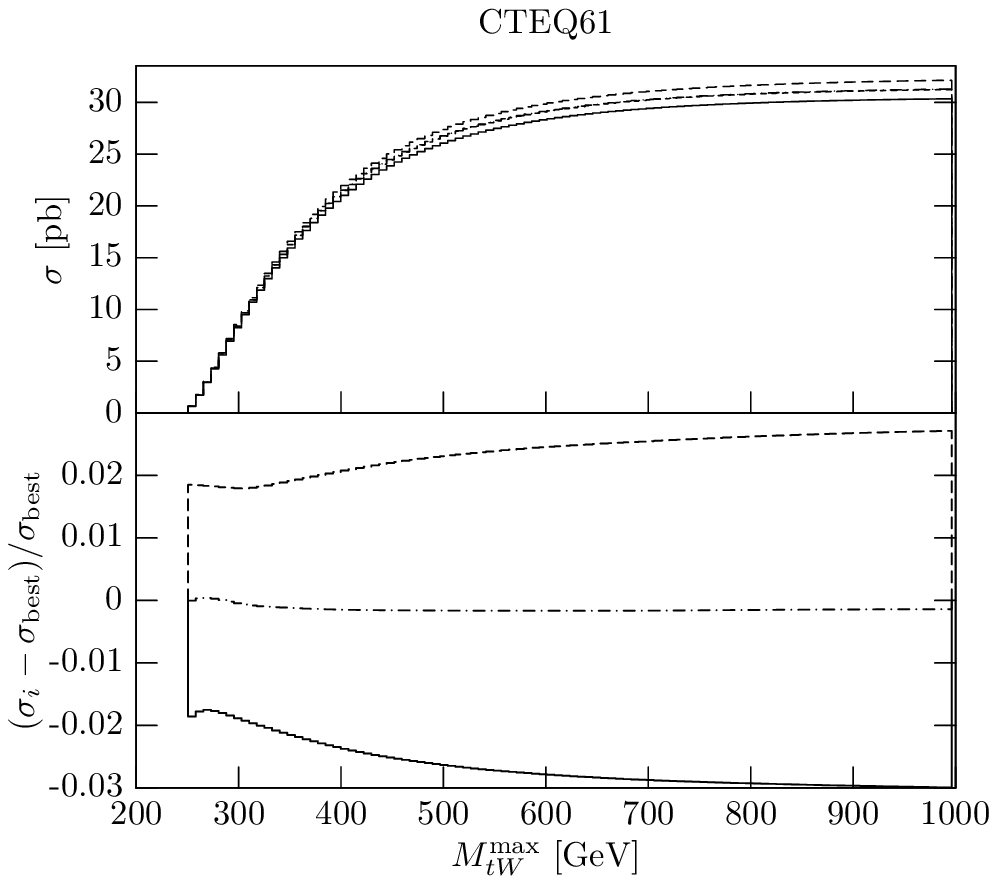,height=8.5cm,width=8cm,angle=0}
\caption{\label{fig:pdfunc} Left: 
Integrated cross section from threshold up to $M_{tW}^{\rm max}$ 
(LO calculation) with the MRST2001E PDF set. For each bin, the minimum 
value (solid line), the maximum (dashed), the average (dot-dashed) and the 
value corresponding to the best fit parton density (dotted) are shown. 
In the lower panel the relative deviation w.r.t. the best fit PDF is shown. 
Right: the same as in the left panel, obtained with the set CTEQ61.}
\end{center}
\end{figure}

\newpage

\begin{figure}
\begin{center}
\epsfig{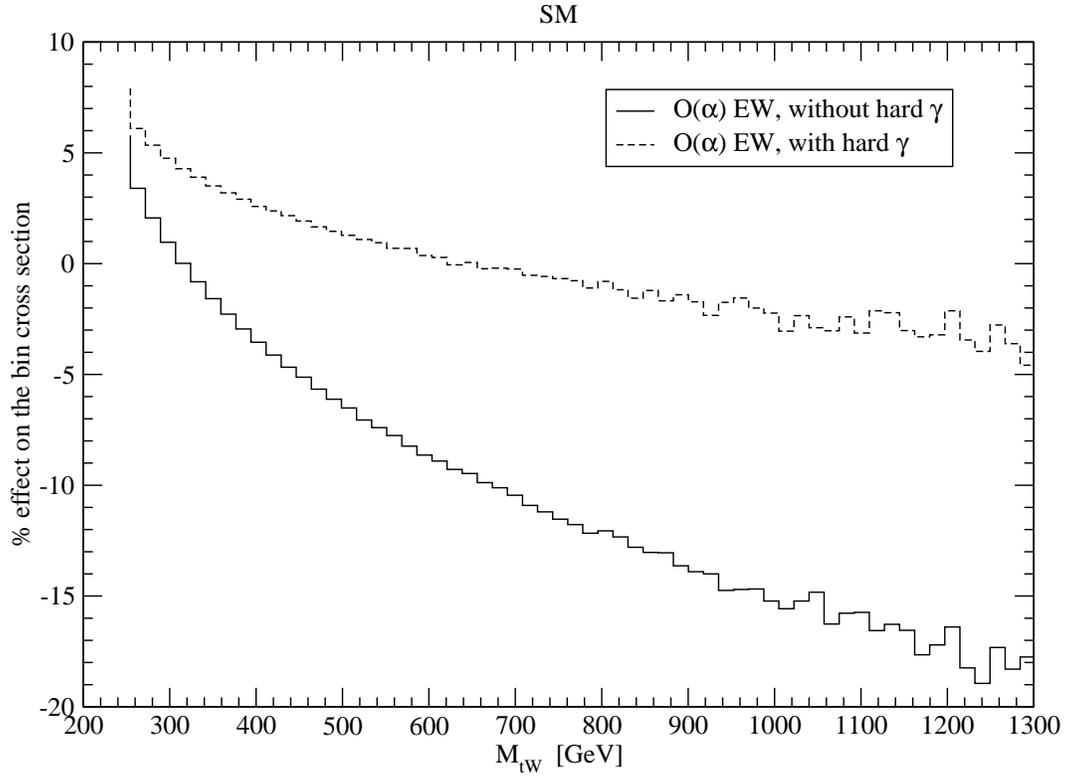}
\caption{
\label{fig:effdistsm} 
Electroweak one-loop effect on the distribution $d\sigma/dM_{tW}$ in the 
Standard Model.
The histogram {\em without hard photon} includes the soft-photon 
contributions only.
}
\end{center}
\end{figure}

\newpage

\begin{figure}
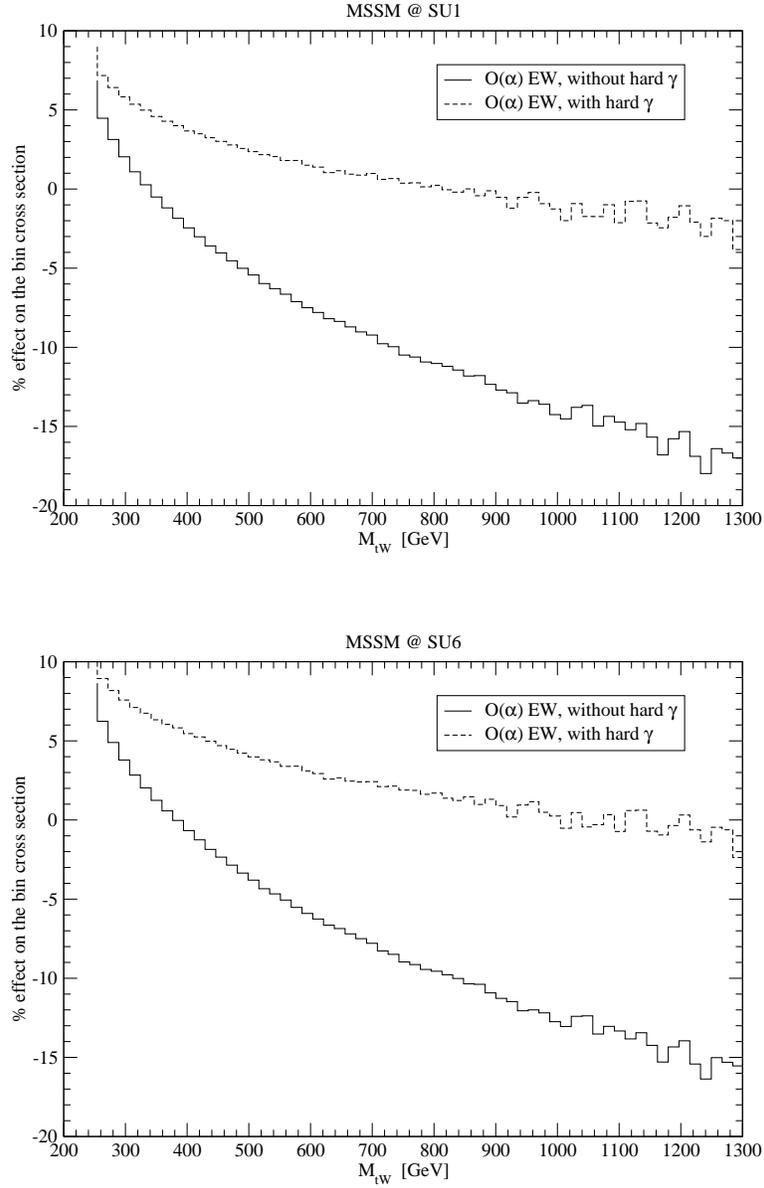

\begin{center}
\epsfig{file=EffectDistributions.SU1.eps,width=10cm,angle=0}
\vskip 1cm
\epsfig{file=EffectDistributions.SU6.eps,width=10cm,angle=0}
\caption{
\label{fig:effdistsusy}
Electroweak one-loop effect on the distribution $d\sigma/dM_{tW}$ in the 
MSSM at the two benchmark points SU1, SU6.
}
\end{center}
\end{figure}

\newpage

\begin{figure}
\begin{center}
\epsfig{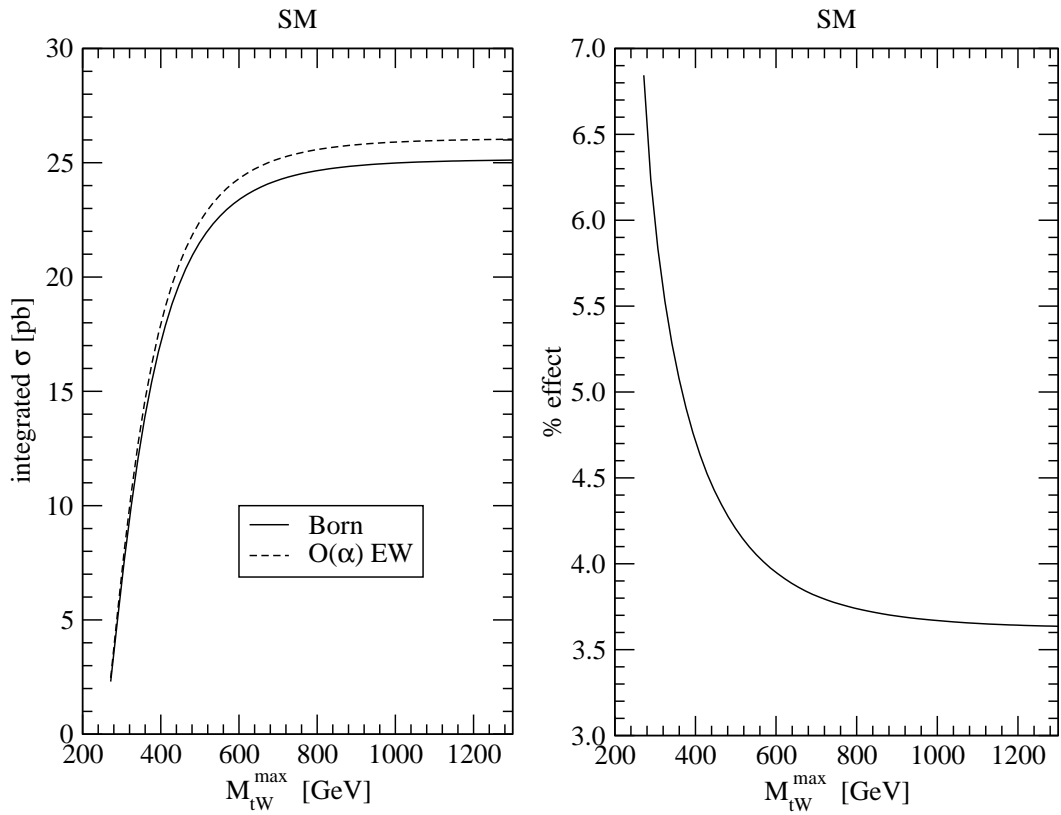}
\vskip 0.8cm
\caption{
\label{fig:effintsm} 
Integrated cross section (from threshold up to $M_{tW}^{\rm max}$) and electroweak one-loop effect in the Standard Model.
}
\end{center}
\end{figure}

\newpage

\begin{figure}
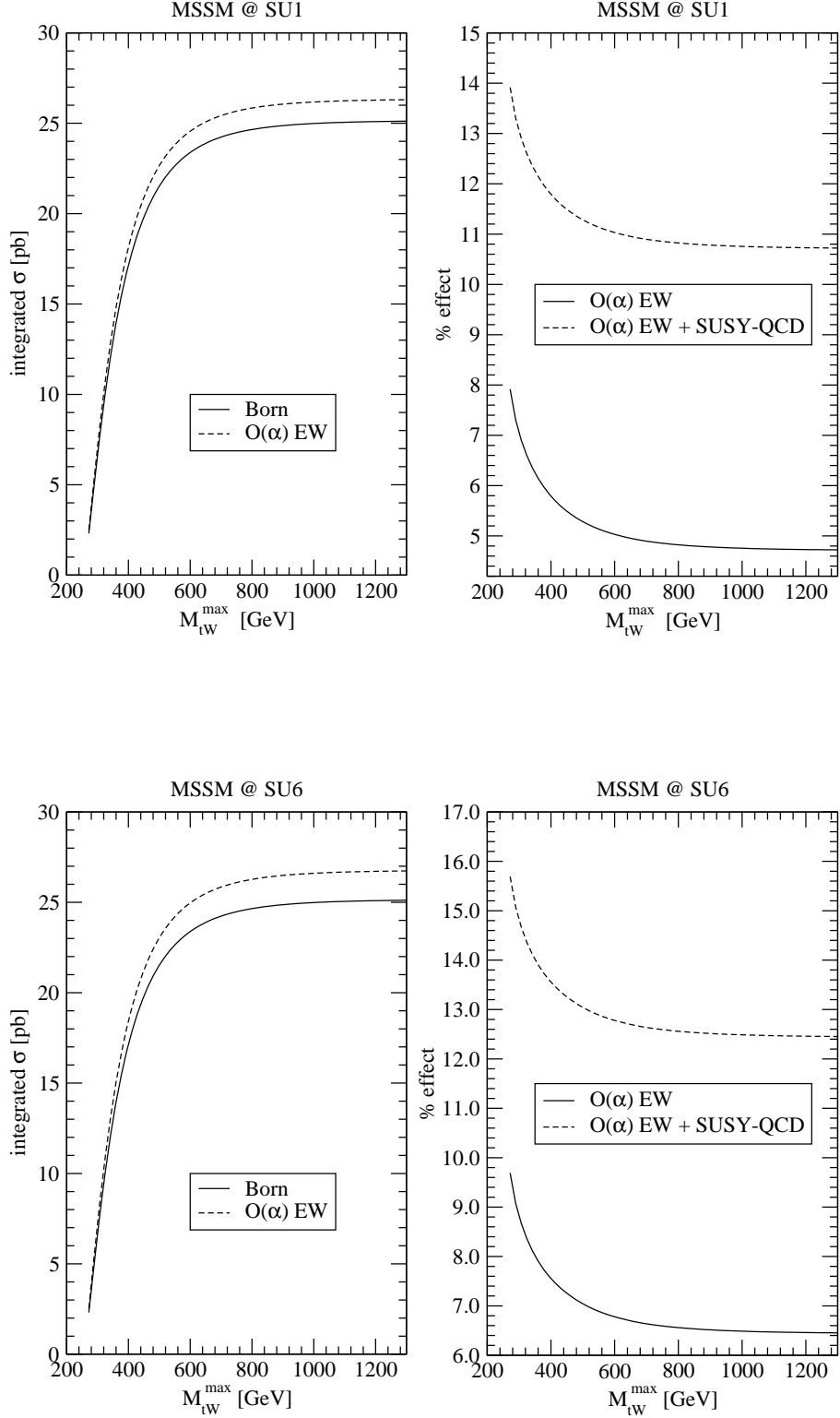

\begin{center}
\epsfig{file=EffectIntegratedSigma.SU1+SQCD.eps,width=12cm,angle=0}
\vskip 2cm
\epsfig{file=EffectIntegratedSigma.SU6+SQCD.eps,width=12cm,angle=0}
\caption{
\label{fig:effintsusy} 
Integrated cross section (from threshold up to $M_{tW}^{\rm max}$) and electroweak one-loop effect in the MSSM at the 
two benchmark points SU1, SU6. SUSY QCD corrections 
have been inserted in the dashed line of right panel and simulated 
by a +6\% shift, consistently with the analysis of Ref.~\cite{Zhang:2006cx}.}
\end{center}
\end{figure}

\end{document}